\documentclass[aip,apl,preprint,superscriptaddress]{revtex4-2}
\usepackage{graphicx}
\usepackage{color}
\usepackage{url}
\usepackage{amsmath} 
 
\usepackage{csquotes}

\begin{document}
\title{Half-Metallic Fe/MgO Superlattice: An Ideal Candidate for Magnetic Tunnel Junction Electrodes}

\author{Nicholas A. Lanzillo$^{a}$}
\affiliation{IBM Research, 257 Fuller Road, Albany, NY 12203, USA}
\thanks{Corresponding Authors (nalanzil@us.ibm.com, apushp@us.ibm.com)} 

\author{Sergey Faleev}
\affiliation{IBM Almaden Research Center, 650 Harry Road, San Jose, CA 95120, USA}

\author{Aakash Pushp$^{a}$,}
\affiliation{IBM Almaden Research Center, 650 Harry Road, San Jose, CA 95120, USA}

%\author{
%	Nick\textsuperscript{1}, 
%	Sergey\textsuperscript{2} and %
%	Aakash\textsuperscript{3} 
%}

%\footnote{Email1}
%\footnote{Email2} 

\date{\today}

\begin{abstract}
Magnetic Tunnel Junction (MTJ) based Spin-Transfer Torque Magnetic Random Access Memory (STT-MRAM) is poised to replace embedded Flash for advanced applications such as automotive microcontroller units. To achieve deeper technological adoption, MTJ needs to exhibit three key features: low magnetization ($M_{s}$), high perpendicular magnetic anisotropy (PMA) and high tunnel magnetoresistance (TMR). Here, we theoretically show that when Fe/MgO multilayers are inserted into the fixed and free layers of the MTJ, these three conditions are simultaneously met. As the number of Fe/MgO multilayers in MTJ electrodes is increased, we find that the electron transport evolves from direct barrier tunneling of majority spin states to the resonant tunneling of minority spin states. Remarkably, the projected density of states (PDOS) of Fe/MgO superlattice at the MgO tunnel barrier exhibits half-metallicity near the Fermi Energy, where the minority states exist while the majority states are gapped out, resulting in astronomically high TMR. 
\end{abstract}

\maketitle

%%%%%%%%%%%%%%%%%%%%%%%%%%%%%%%%%%%%%%%%%%%%%%%%%%%%%%%%%%%%%%%%%%%%%%%%

\section{Introduction}
Magnetic Random Access Memory (MRAM) is a special class of electronic memory that relies on the spin of the electron rather than its charge – thus it is dubbed a spintronic device. Owing to its electronic nature, it offers unlimited endurance, and owing to its persistent spin based magnetic nature, it offers non-volatility. So, in essence, it has all the key attributes for an ideal memory element, which justifies tremendous continued academic and industrial interest over the past $\approx$50 years (since 1975\cite{worledge2024spin}.)

MRAM is a 2-Dimensional crossbar array of Magnetic Tunnel Junctions (MTJs) that are formed from two ferromagnets sandwiching a tunnel barrier. When the two ferromagnetic electrodes are magnetized parallel to each other the tunnel resistance ($R_{P}$) across the MTJ is low, and when they are magnetized anti-parallel to one another the tunnel resistance ($R_{AP}$) is high. The electrical resistance is measured by passing a small current across the MTJ, and switching of the magnetic orientation of one of the layers (called the free layer, FL) is achieved by flowing a higher current beyond a threshold. By switching the polarity of the threshold current, the FL can be magnetically switched back and forth thereby accessing the two resistance states forming a two-terminal binary memory element. This makes the so-called spin transfer torque (STT) - MRAM extremely promising. To ensure that only one of the two ferromagnetic layers switches, we form the non-switchable magnetic electrode, called the Reference Layer (RL) from a synthetic anti-ferromagnet (SAF), which is formed from two ferromagnetic layers sandwiched by spacer metal layer that provides anti-ferromagnetic interlayer exchange coupling\cite{parkin1990oscillations,worledge2024spin}.

Typically, MTJ electrodes are formed from CoFeB alloys\cite{worledge2024spin} and more recently from Heusler Alloys\cite{garg2025ferri} with MgO being the best tunnel barrier\cite{parkin2004giant,yuasa2004giant} to obtain ON/OFF ($R_{AP}/R_{P}$) ratio of 2-3x, corresponding to Tunneling Magneto Resistance TMR: $100*(R_{AP}-R_{P})/R_{P}$ $\approx 100-200\%$ at ambient temperature. STT-MRAM is set to replace eFLASH beyond 28nm node in all advanced foundries, and is planned to be used in next generation automotive microcontroller units\cite{shih2020reflow}. STT-MRAM is at the cusp of mainstream high-performance memory (e.g. last-level cache) adoption, once these three key attributes are simultaneously satisfied: \newline
1. The FL should have low magnetic moment ($M_{s}$) for low switching current. \newline
2. The FL should have high Perpendicular Magnetic Anisotropy (PMA) for bit stability. \newline
3. The TMR should be as high as possible. \newline 
If the TMR were higher, it would allow for MTJs of smaller critical dimensions ($CD < 30nm$) thereby decreasing the switching current, improving switching efficiency and reducing write error rates (WERs\cite{kent2015new}.)

In this theoretical work, we show that MTJ electrodes formed from Fe/MgO superlattices meet all these three criteria. Additionally, owing to high phonon energy of the ceramic MgO layers infused in between Fe layers, we speculate that electron-boson (-phonon/-magnon) coupling in Fe/MgO superlattice based FL will be reduced thereby leading to low Gilbert damping – a critical requirement for ultrafast switching.

It is well known in the spintronic community that Fe/MgO interface offers PMA\cite{klaua2001growth,dieny2017perpendicular,monso2002crossover,worledge2011spin}. We extended this understanding and recently proposed to form the magnetic electrodes from “bulk” of such Fe-MgO interfaces, i.e., Fe/MgO superlattice\cite{pushp2019magnetic}. We found that the PMA for such an electrode is exceptionally high comparable to Ferrimagnetic Heusler Alloys\cite{faleev2017origin}, while the $M_{s}$ is reduced due to the incorporation of MgO sublayers. Remarkably, we find that the PDOS of Fe/MgO superlattice at the MgO tunnel barrier behaves like a half-metal, where only the minority states exist, while the majority states are gapped out, resulting in astronomically high TMR. In the extreme case of the Fe/MgO superlattice on one side of the MgO tunnel barrier and bulk Fe on the other side, we obtain negligible TMR signifying the distinct electron transport derived from the minority state PDOS of the Fe-MgO superlattice. In the following, we discuss these key findings.

\section{Methods} 
In order to calculate the magnetic anisotropy of various Fe/MgO multilayer structures we performed density functional theory (DFT) calculations for n[Fe]/m[MgO] multilayers (periodically repeated along the z-axis - the direction normal to the interface plane) with different numbers of Fe atomic layers, n, and MgO atomic layers, m. We used the generalized gradient approximation (GGA) implemented in the VASP program\cite{kresse1996efficient} with projector augmented wave potentials\cite{blochl1994projector,kresse1999from} and the PBE GGA/DFT functional\cite{perdew1997generalized}. 

The volume magnetocrystalline anisotropy (MCA) energy per formula unit, $K_{mc}$, of the n[Fe]/m[MgO] multilayer is calculated as the difference between the total energies of the magnetic states with magnetization directed along the x-axis and the z-axis, i.e. $K_{mc} = E_{100} - E_{001}$, where positive $K_{mc}$ corresponds to out-of-plane magnetization. For MCA calculations, we used finer $24x24xn_{z}$ mesh in reciprocal space (with corresponding $n_{Z}$ inversely proportional to the thickness of the multilayer in the z-direction) and included spin-orbit interactions self-consistently in the DFT cycle. We also calculated the volume magnetic anisotropy, $K_{v} = K_{mc} - K_{sh}$, where $K_{sh} = \frac{1}{2} u_{0} M_{s}^2 V$  is the shape anisotropy energy of a thin film per unit cell volume V, $M_{s}$ is the saturation magnetization and $u_0$ is the vacuum permeability. 

All electron transport calculations were performed using DFT and Non-Equilibrium Green's Function (NEGF) formalism as implemented in the QuantumATK software suite\cite{quantumatk}. We used the scalar-relativistic SG15 pseudopotential\cite{schlipf2015sg15}, a numerical atom-centered basis set and a cutoff energy of 150.0 Hartree for all elements. For the exchange-correlation functional, we employed the generalized-gradient approximation\cite{perdew1997generalized}. Electron transport calculations using the NEGF formalism were carried out using a two-terminal setup with semi-infinite left/right electrodes and a finite scattering region in the middle. Lateral k-point sampling of $45x45$ was used for all transport calculations with dense sampling of 201 in the z-direction for the semi-infinite electrode regions. Transport calculations have semi-infinite Fe electrodes for all structures.  

We consider the face-centered cubic (FCC) phase of MgO with a lattice constant of 4.09\AA{} while iron is taken to be body-centered cubic (BCC) with an equilibrium lattice constant of 2.86\AA{}. The lateral lattice constants in the directions perpendicular to transport were matched to the lattice constant of Fe. 

\section{Results}
We first consider periodic stacks of bulk Fe/MgO multilayers as described above. An image of the supercell corresponding to two atomic layers Fe and two atomic layers of MgO is shown in Fig. 1(a). The calculated total energy as a function of magnetic moment orientation is shown in Fig. 1(b), where we see that the out-of-plane energy is lower than the in-plane energy, thereby confirming this Fe/MgO multilayer stack possesses perpendicular magnetic anisotropy (PMA). These results are in agreement with previous reports in the literature for magnetic ansitropy at the Fe/MgO interface\cite{koo2013large,lambert2023quantifying}.
\begin{figure*}[h!]
	\centering%
	\begin{center}
		\includegraphics[scale=0.5,clip,trim = 0mm 0mm 0mm 0mm]{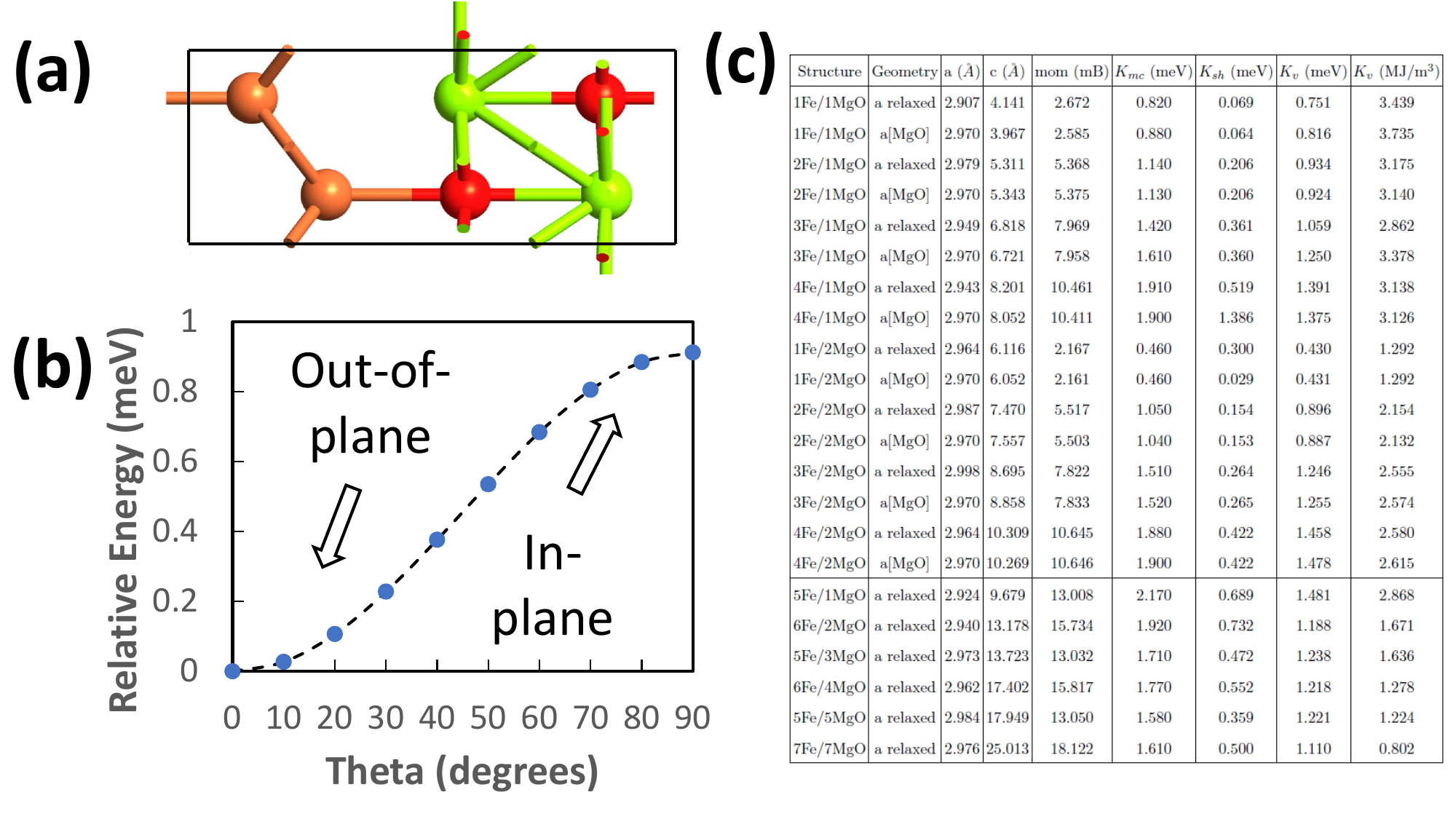}
	\end{center}
	\caption {(a) Atomic scale representation of a unit cell consisting of two atomic layers of Fe and two atomic layers of MgO. Both layers are oriented such that the z-axis points along the (100) direction (b) Calculated total energy of the system as a function of the magnetic polarization angle relative to the normal (z-axis). (c) A table with the calculated magnetic anisotropy for various Fe/MgO multilayer configurations. The in-plane lattice constant of multilayer structures n[Fe]/m[MgO], a, out-of-plane lattice constant, c, total magnetic moment, $mom$, the MCA energy, $K_{mc}$, the shape anisotropy energy, $K_{sh}$, the volume magnetic anisotropy, $K_{v}$ (in units of meV per formula unit, perpendicular magnetic anisotropy corresponds to positive $K_{v}$). The top part of the table contains smaller multilayers with the number of Fe atomic layers n=1,2,3,4 and the bottom part of the table contains the multilayers with n=5,6,7. For multilayers with n=1,2,3,4, we present the results for both relaxed in-plane lattice constant (indicated as \enquote{a relaxed}) and in-plane lattice constant fixed to that of the MgO (indicated as \enquote{a[MgO]}).}
\end{figure*}
This PMA is preserved as the number of Fe and/or MgO layers varies. The calculated values of MCA for various Fe/MgO multilayer structures are summarized in Fig. 1(c).

As one can see from Fig. 1(c), all considered n[Fe]/m[MgO] multilayers have perpendicular magnetic anisotropy (PMA) with volume magnetic anisotropy $K_{v} > 1$ meV for all considered multilayers (except 7Fe/7MgO). The $K_{v}$ increases for shorter multilayers that have larger Fe/MgO interface area per unit length in the z-direction, which is expected since the anisotropy in the multilayers originates from the Fe/MgO interface. 

For multilayers with the number of Fe atomic layers n=1,2,3,4 we present results for both relaxed (obtained by total energy minimization) in-plane lattice constant (indicated as "a relaxed" in Fig. 1(c)) and in-plane lattice constant fixed to that of MgO (indicated as "a[MgO]" in Fig.1 (c).) In the "a[MgO]" case we used an MgO lattice constant of 4.20 A that corresponds to $a=\frac{4.20}{\sqrt(2)} = 2.97$A. The out-of-plane lattice constant, c, is always relaxed (obtained by energy minimization.) One can see that the results for the magnetic moments, $K_{v}, K_{mc}$ and $K_{sh}$ are close for both lattice constants for most of the considered multilayers. 

For representative MTJ structures, we consider a 6-atomic-layer thick MgO tunnel barrier separating the reference layer on the left-hand side of the device from the free layer on the right-hand side of the device. The fixed and free layers are composed of Fe(100). We then insert $N=M={0,1,2,3}$ Fe/MgO multilayers on either side of the tunnel barrier, with $N=M=0$ corresponding to the reference structure with fixed/free layers composed entirely of Fe. These structures are depicted in Fig.2(a). As a limiting case, we also consider the case of $N=M=\infty$ Fe/MgO multilayers, where the semi-infinite electrodes are composed of Fe/MgO instead of Fe, as depicted in Fig. 2(b). 
%\begin{figure}[h!]%
%	\centering%
%	\begin{center}
%		\includegraphics[scale=0.25,clip,trim = 0mm 0mm 0mm 0mm]{Fig2}
%	\end{center}
%	\caption {Schematic showing the two-terminal set up used for NEGF transport calculations.}
%\end{figure}
Relative to the pristine Fe/MgO/Fe MTJ structure, the presence of Fe/MgO multilayers in the fixed and free layers results in non-trivial changes to the electronic structure. For the limiting cases of MTJ structures with $N=M=0$ Fe/MgO multilayers (a reference Fe/MgO/Fe MTJ) and $N=M=\infty$ (the structure in Fig. 2(b)) we plot the spin-resolved projected density of states (PDOS) in Fig. 2(c) and (d), respectively. 
\begin{figure}[h!]
	\centering%
	\begin{center}
		\includegraphics[scale=0.5,clip,trim = 0mm 0mm 0mm 0mm]{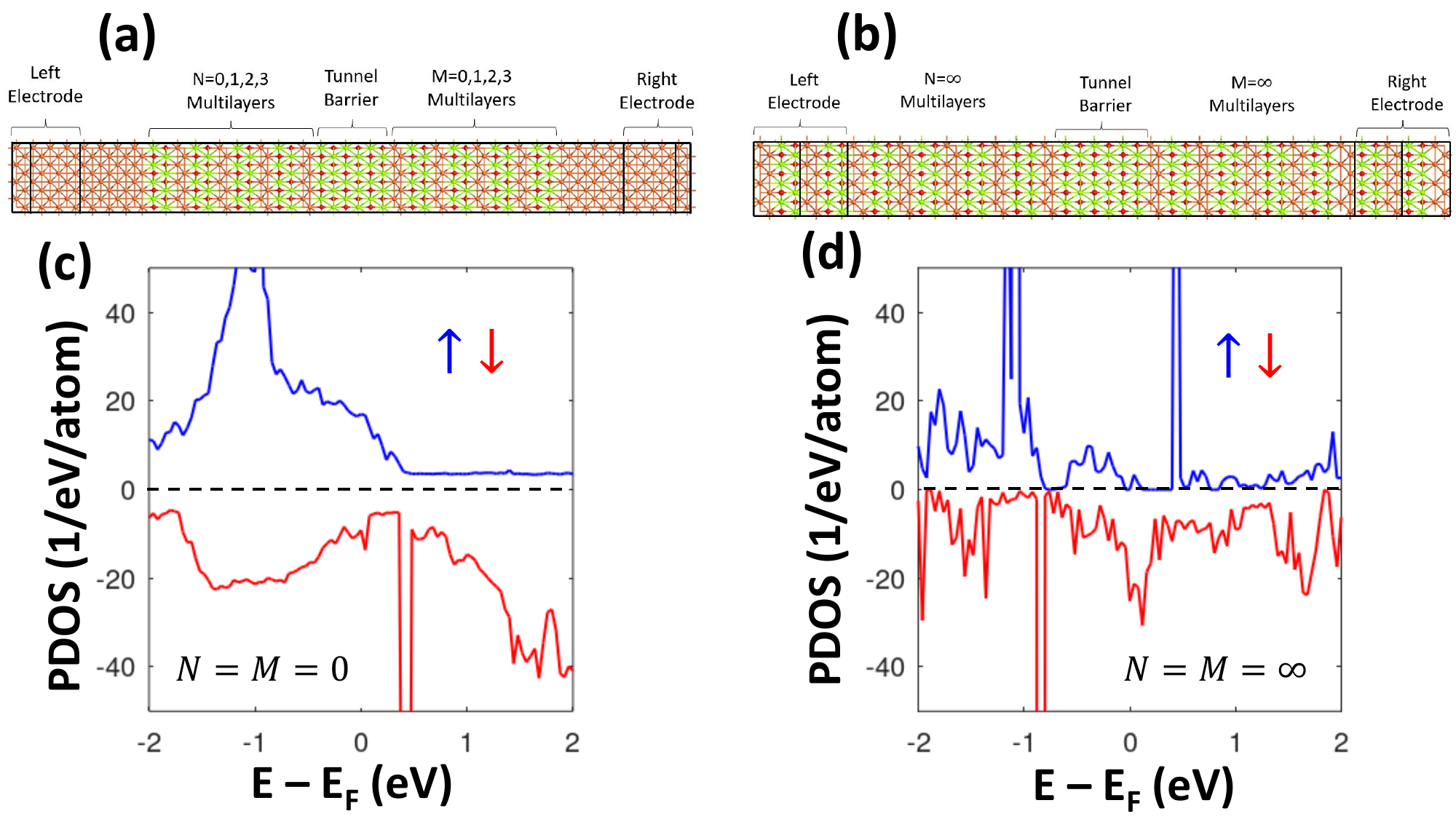}
	\end{center}
	\caption {(a) A magnetic tunnel junction with N Fe/MgO multilayers in the reference layer and M Fe/MgO layers in the free layer; both having semi-infinite Fe electrodes (b) An MTJ with semi-finite Fe/MgO electrodes for both reference and free layers (c) Projected device density of states (PDOS) for the reference MTJ (N=M=0; Fig. 2(a)) and the MTJ composed entirely of Fe/MgO multilayers ($N=M=\infty$, Fig. 2(b).)}
\end{figure}
Since states are occupied up to the Fermi Energy ($E_F = 0$ here) we see that the occupation in the reference structure is dominated by spin-up electrons. This is true both at the Fermi Energy itself as well as for all energies less than $E_F$. In the unoccupied energy window ($>E_F$) we see dominant contributions from the spin-down electrons, including a large peak between 0 eV and 1 eV which stems from the Fe atoms close to the Fe/MgO interface. For the Fe/MgO multilayer structure, we see an inversion of the dominant PDOS in Fig. 2(d). The spin-down electrons have a larger PDOS at $E_F = 0$ than the spin-up electrons. In fact, near the Fermi Energy, the spin-up PDOS at the tunnel barrier is gapped out. The large unoccupied peak in the spin down channel between 0 eV and 1 eV in the reference structure becomes occupied in the Fe/MgO mutlilayer structure. An additional unoccupied peak appears in the spin-up channel. 

In order to better understand the dependence of the PDOS on the structural characteristics of the MTJs, we plot the local density of states (LDOS) as a function of device length in Fig. 3. 
\begin{figure}[h!]
	\centering%
	\begin{center}
		\includegraphics[scale=0.6,clip,trim = 0mm 0mm 0mm 0mm]{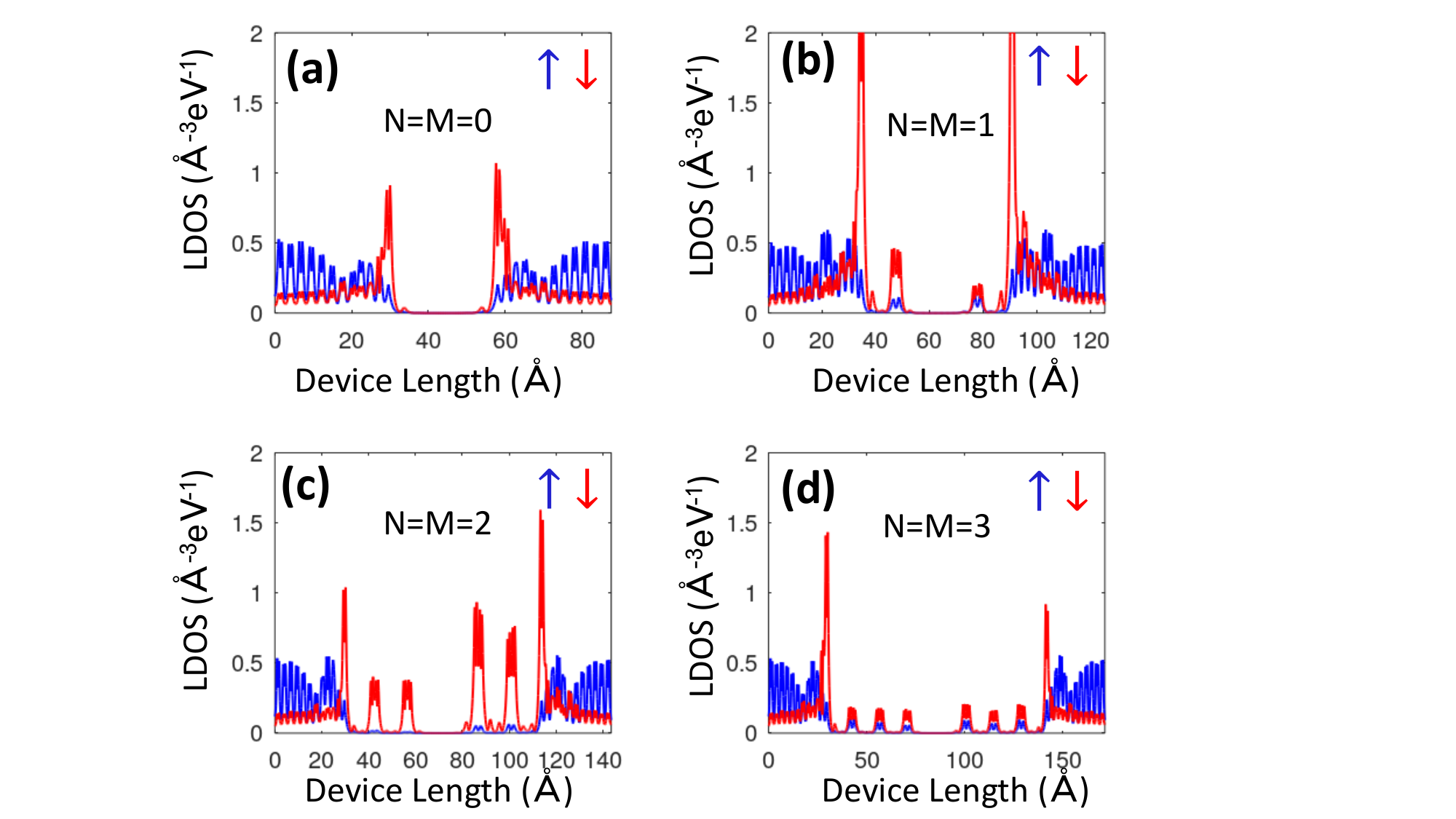}
	\end{center}
	\caption {Local density of states (LDOS) as a function of device length for $N=M={0,1,2,3}$ Fe/MgO multilayers on either side of a 6-atomic-layer thick MgO barrier in the spin-parallel configuration. Blue curves are for spin up electrons and red curves are for spin down electrons.}
\end{figure}
For a standard Fe/MgO/Fe MTJ in the spin-parallel state, electronic density of states in the Fe electrodes is dominated by the spin-up (majority) electron contribution while localized peaks of the spin-down electrodes at the two Fe/MgO interfaces\cite{butler2001spin}. This is plotted in Fig. 3(a), which we refer to as the $N=M=0$ case. The localized minority spin peak at the Fe/MgO interface is the interface resonant state which has been reported previously in the literature\cite{butler2001spin,duluard2015enhanced}. As the number of Fe/MgO multilayers increases from $N=M=0$ to $N=M=3$, the local density of states (LDOS) in the electrode regions remains largely unchanged relative to the ordinary Fe/MgO/Fe case: we see that spin-up dominates with localized peaks from spin-down at the edge of the electrodes where the Fe/MgO interface is present. However, we see peaks emerge in the Fe/MgO multilayer regions which are dominated by the spin-down electrons while the spin-up contribution is strongly suppressed. Since the Fe/MgO multilayers create additional Fe/MgO interfaces, they provide additional localization points for minority spin carriers. The presence of these peaks has a non-trivial impact on electron transport across these MTJ structures. 

As has been reported previously, the electron transmission spectrum for the reference Fe/MgO/Fe MTJ structure is dominated by a symmetric peak in reciprocal space for the majority spin carriers (spin-up) centered around the $\Gamma$-point ($k_x = k_y = 0$) which is indicative of direct barrier tunneling\cite{butler2001spin}. The minority spin (spin-down) shows smaller peaks throughout the Brillouin Zone (BZ) which are indicative of resonant tunneling. These trends are reproduced by our results as seen in Fig. 4(a) for the reference structure with $N=M=0$. As the number of multilayers is increased, the symmetric peak at $\Gamma$ as well as the resonant tunneling peaks become smaller, reflecting the reduced overall transmission due to presence of more MgO layers in the structure. However, we notice that the heights of the resonant tunneling peaks gradually become larger than the height of the direct tunneling peak with increasing number of multilayers. By the time $N=M=3$, the direct barrier tunneling peak is almost completely suppressed and the transport becomes dominated by the resonant tunneling channels of the minority spin carriers, as shown in Fig. 4(d). 
\begin{figure}[h!]
	\centering%
	\begin{center}
		\includegraphics[scale=0.5,clip,trim = 0mm 0mm 0mm 0mm]{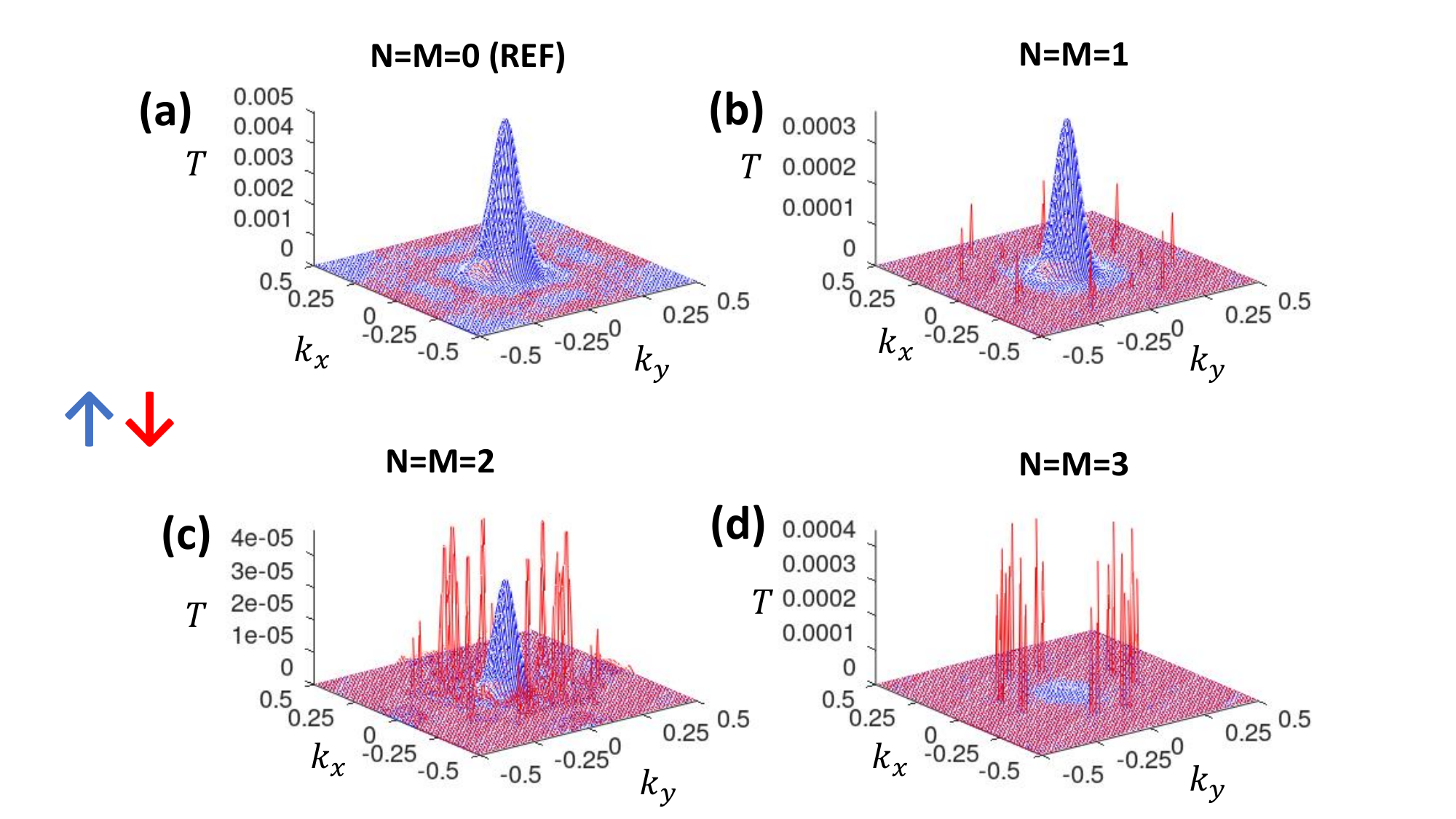}
	\end{center}
	\caption {Reciprocal-space-resolved transmission spectra for magnetic tunnel junctions with $N=M={0,1,2,3}$ Fe/MgO multilayers on either side of the tunnel barrier. Blue corresponds to the majority spin (up) and red corresponds to the minority spin (down.)}
\end{figure}
The inversion observed between direct barrier tunneling and resonant tunneling with increasing number of Fe/MgO multilayers has a significant impact on total transmission and TMR. In Fig. 5(a) we plot the total transmission for both majority (spin up) and minority (spin down) carriers as a function of N(M). For the parallel configurations, we see that $N=M=0$ has a larger contribution from the majority spin than the minority spin (up-up v. down-down) by the time $N=M$ is increased to 3 the minority spin has the largest contribution. The transmission in the anti-parallel state (which is equal for majority and minority spin due to time-inversion symmetry) shows a continuous and significant reduction as N is increased, dropping by approximately five orders of magnitude. 
\begin{figure}[h!]
	\centering%
	\begin{center}
		\includegraphics[scale=0.5,clip,trim = 0mm 0mm 0mm 0mm]{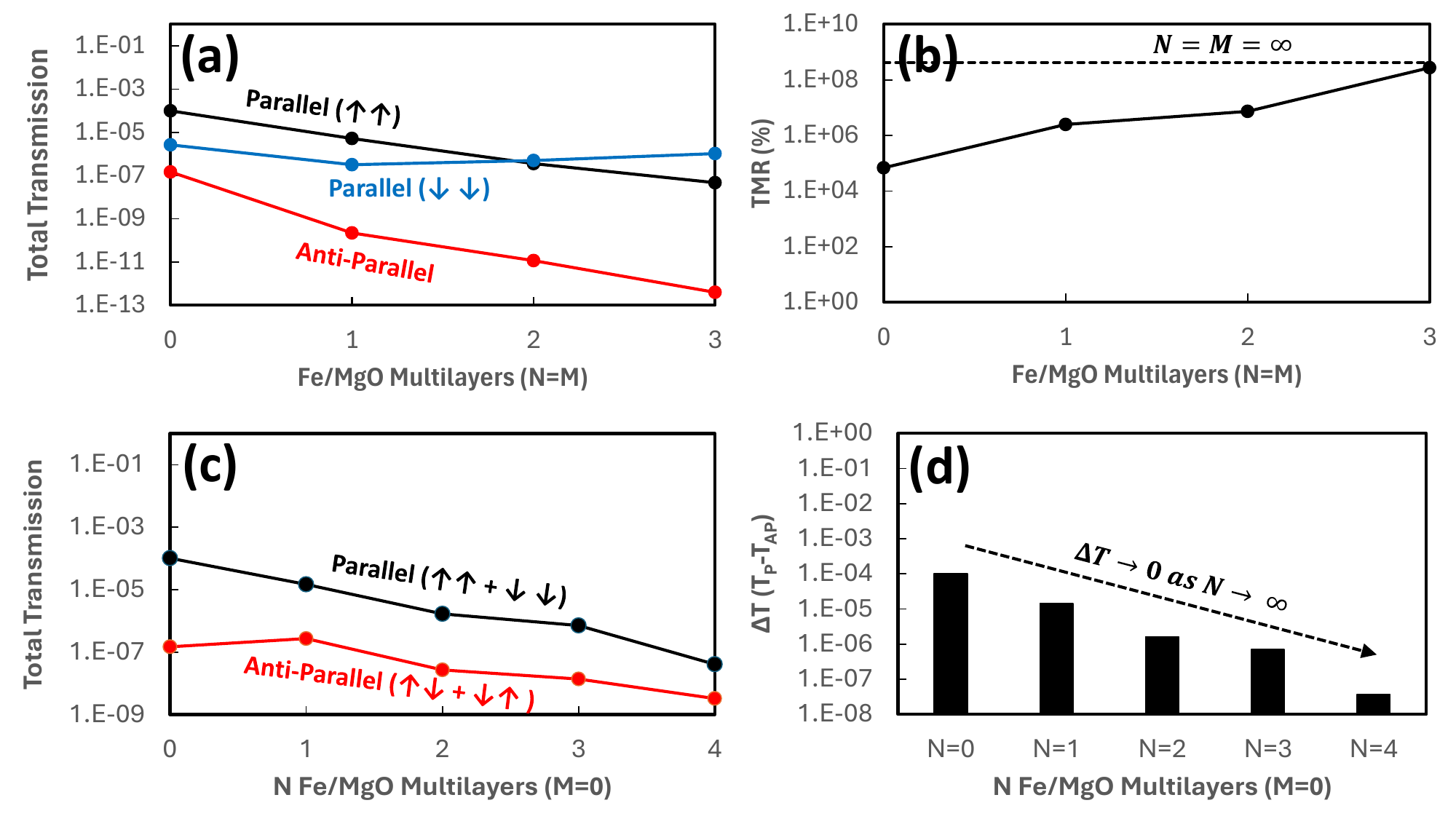}
	\end{center}
	\caption {(a) Total electron transmission for parallel (up,up), parallel (down,down) and anti-parallel MTJ configurations with $N=M={0,1,2,3}$ layers of Fe/MgO. (b) Calculated TMR for the same structures. (c) total electron  transmission and (d) difference between parallel and anti-parallel transmission for MTJ configurations with M=0 on one side and $N={1,2,3,4}$ on the other side.}
\end{figure}
Since TMR depends on the difference between transmission in the parallel v. anti-parallel states, the strong suppression of anti-parallel spin transport coupled with a weaker modulation of the parallel spin transport results in a significant increase in TMR as plotted in Fig. 4(b). The TMR increases by approximately three orders of magnitude, from $\approx 10^5$ for the reference structure to $\approx 10^8$ for $N=M=3$. The calculated TMR for the $N=M=3$ case is within a few percent of the calculated value for $N=M=\infty$, suggesting that the transmission trends with respect to N saturate after just a few multilayers are added.
Interestingly, if we construct MTJs with Fe electrodes on one side (M=0) and N Fe/MgO multilayers on the other side, we see that the total transmission for both parallel and antiparllel configurations rapidly decreases with increasing number of multilayers, as depicted in Fig.5(c). The corresponding difference between the total parallel and anti-parallel transmission values is plotted in Fig. (d), where we observe roughly an order-of-magnitude decrease with each additional Fe/MgO mutlilayer. In the limit of $N \rightarrow \infty$ (with M=0) the difference in transmission approaches zero. Identically zero transmission for parallel and anti-parallel configurations by definition results in zero TMR. 

\section{Conclusion and Experimental Outlook}
Having discussed the theoretical aspects of Fe/MgO superlattice based MTJ electrodes, we briefly touch upon the experimental feasibility of our proposal. Fe/MgO interfaces\cite{li1991giant}, Fe/MgO/Fe sandwiches\cite{klaua2001growth} and discontinuous metal-insulator multilayers (DMIM)\cite{vovk2019probing} have been intensely studied both theoretically and experimentally. The most interesting experimental work corroborating our approach is from Fahsold et al.\cite{fahsold2000growth}, who have experimentally demonstrated that growing Fe on MgO below 140K allows for mitigating wetting issues of Fe on MgO owing to surface energy difference. Growing MgO on top of ultra thin Fe underlayers should be rather straightforward as has been experimentally demonstrated by numerous papers\cite{klaua2001growth,martinez2003epitaxy,martinez2005coverage,vovk2019probing}. We speculate that growing the first layer of Fe on MgO at cold temperatures to mitigate wetting issues, followed by growing subsequent Fe layers at elevated temperatures needs to be carefully explored to obtain Fe/MgO superlattices with continuous Fe sublayers considered in our study.

If our theoretical prediction is experimentally realized, where we obtain TMR of $\approx500-1000\%$ at ambient temperature with low $M_{s}$ and high PMA of the FL, not only will it immediately impact STT-MRAM but will also enable Voltage Controlled Magnetic Anisotropy (VCMA)-MRAM\cite{nozaki2019recent}, multi-bit MRAM\cite{pushp2022magnetic} for neuromorphic applications, and in-memory computing approaches for multiply accumulate macros.

\textbf{Acknowledgements:} We gratefully acknowledge discussions with B. Madon, M. A. Mueed, S. Rebec, N. Arellano, T. Topuria, E. Delenia, A. Fong, S. Narayan, B. Kurdi, W. Wilcke, T.C. Chen, M. Ritter, V. Narayanan and D. C. Worledge.

%\bibliography{mram} 

\end{document}